\documentclass[twocolumn]{article}

\usepackage[backend=biber,style=nature]{biblatex}
\usepackage[acronym,nomain]{glossaries}
\glsdisablehyper
\newacronym{eqprop}{EqProp}{equilibrium propagation}
\newacronym{backprop}{BackProp}{backpropagation}
\newacronym{bptt}{BPTT}{backpropagation through time}
\newacronym{pbc}{PBC}{periodic boundary conditions}

\usepackage{graphicx}
\usepackage{hyperref}
\usepackage{cleveref}
\usepackage{dsfont}
\usepackage{float}
\usepackage{bm}
\usepackage{subfig}
\usepackage{pifont}
\usepackage{dutchcal}
\usepackage[normalem]{ulem}
\usepackage{multirow}
\usepackage{amsmath}
\usepackage{amssymb}
\usepackage[labelfont=bf,font=footnotesize]{caption}
\usepackage[textwidth=7.25in, textheight=9.5in]{geometry}
\usepackage[percent]{overpic}
\usepackage{xcolor}
\usepackage{amsfonts}
\usepackage{textcomp}
\usepackage{nth}
\usepackage{siunitx}
\usepackage{verbatim}

\newcommand\D{\mathrm d}
\newcommand\E{\mathrm e}

\renewcommand{\vector}[1]{{\bm{#1}}}

\newcommand{\period}{\tau}

\newcommand{\refappendix}[1]{Appendix~\ref{#1}}

\newcommand{\mybox}[1]{\textbf{\Large{(#1)}}}
\newcommand{\phantomsubfloat}[1]{
        {% apply caption setup only temporarily
        \captionsetup[subfigure]{labelformat=empty}
        \subfloat[][]{#1}
    }%
}

\title{Equilibrium Propagation for Dissipative Dynamics}% Force line breaks with \\

\usepackage{authblk}
\author{Marc Berneman}
\author{Daniel Hexner}
\affil{Faculty of Mechanical Engineering, Technion - Israel Institute of Technology, 3200003 Haifa, Israel}
\affil{\texttt{marcberneman@campus.technion.ac.il}, \texttt{danielhe@technion.ac.il}}

\begin{document}

    \twocolumn[
        \begin{@twocolumnfalse}
            \maketitle
            \begin{abstract}
                \noindent
                Computing gradients of a cost function is central to design-based optimization and machine learning algorithms.
                Equilibrium propagation provides an exact method to compute gradients in hardware by exploiting the inherent physical laws.
                The locality of these algorithms, in conjunction with local updates, enables mechanical and electronic systems that autonomously learn a function.
                We extend these methods to damped dynamical systems operating in the linear regime, such as mechanical structures obeying damped Newtonian dynamics and RLC circuits.
                By introducing an effective action whose extremum corresponds to the underlying dynamics, we derive local learning rules.
                This approach applies both to problems with periodic boundary conditions and to those with resting initial conditions.
                We demonstrate the viability of our method in mechanical and electronic systems and explore novel functionality such as classifying temporal sound signals.
                Our work opens the door to intelligent materials that process dynamical signals, enabling temporal computations, passive and active sensors, and materials that act as frequency-dependent filters.

            \end{abstract}

        \end{@twocolumnfalse}
    ]

    The fusion of ideas from learning with the physics of mechanical and flow systems provides a platform for material and material-like systems to learn~\cite{scellier2017equilibrium,stern2021supervised,hexner2020periodic,vadlamani2020physics, lee2022mechanical,anisetti2023learning,falk2023learning,patil2023self,stern2023learning,du2025metamaterials}.
    From the materials science perspective, an ordinary material can acquire complex machine-like functionality, without design nor intervention with the microscale~\cite{pashine2019directed,hexner2020periodic,stern2020supervised,stern2021supervised,anisetti2023learning}. From the computing perspective, this offers a decentralized, efficient, and reconfigurable platform for sensors, filters, and processors, with integrated learning degrees of freedom~\cite{dillavou2022demonstration,momeni2024training}.

    At the heart of physical learning are algorithms that exploit physical laws to compute the gradient of a cost function.
    Contrastive learning~\cite{ackley1985learning,movellan1991contrastive,stern2021supervised,stern2024physical,stern2025physical} and \gls{eqprop}~\cite{scellier2017equilibrium} provide exact methods for gradient calculations using only two measurements.
    The only requirement is that the state of the system extremizes an energy function, which need not be the actual physical energy.
    To date, most work dealing with physical systems has focused on systems operating quasistatically.
    Examples include resistor networks~\cite{kendall2020training,dillavou2022demonstration,dillavou2022demonstration,stern2022physical,dillavou2024machine,anisetti2024frequency}, flow networks~\cite{stern2021supervised}, and elastic networks~\cite{stern2021supervised,altman2024experimental}.

    Generalizing \gls{eqprop} to the dynamical regime is challenging, since the trajectory typically does not extremize an energy function~\cite{lopez2023self,mandal2024learning,pourcel2025learning}.
    There have been proposals that Lagrangian systems can be trained by leveraging the principle of least action~\cite{massar2025equilibrium,pourcel2025lagrangian}.
    Approximate schemes have been proposed for dissipative systems~\cite{massar2025equilibrium}, and certain approaches exploit fixed boundary conditions~\cite{kendall2021gradient}.
    However, these remain limited in scope, and a general framework for learning in dissipative dynamical regimes has yet to be established.

    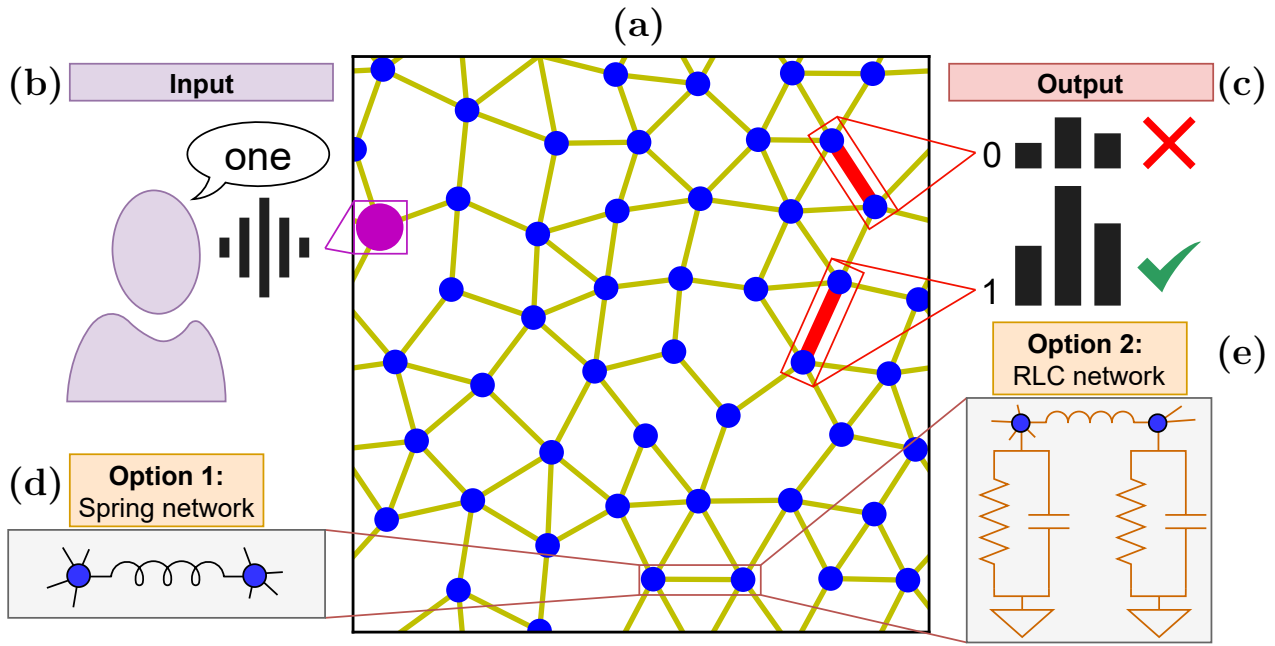
\begin{figure*}[t]
        \centering
        \subfloat{
            \begin{overpic}{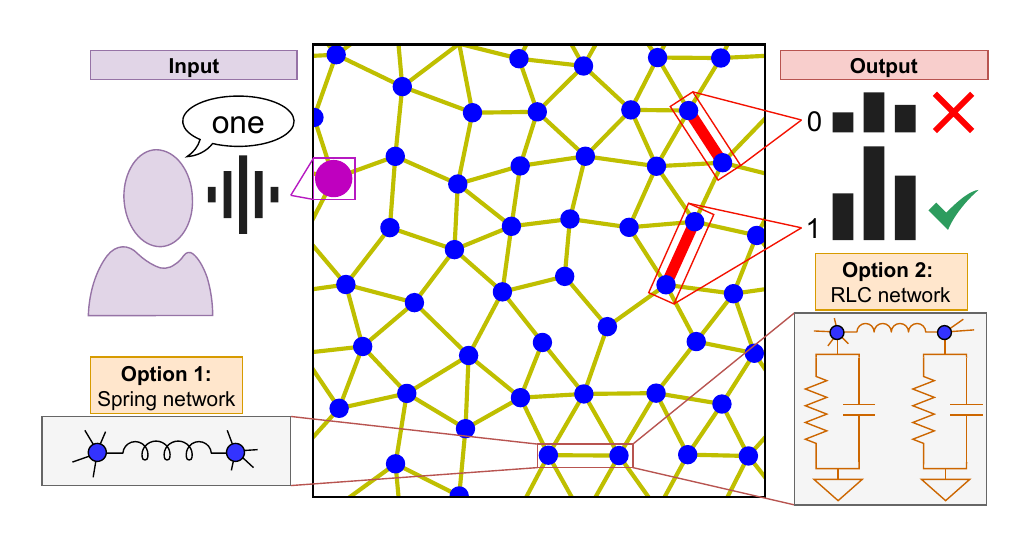}
                \put(50,50){\mybox{a}}
                \put(4,45.5){\mybox{b}}
                \put(96,45.5){\mybox{c}}
                \put(4,15){\mybox{d}}
                \put(96,25){\mybox{e}}
            \end{overpic}
            \label{fig:intro_figure_net}
        }\\
        \phantomsubfloat{\label{fig:intro_figure_input}}
        \phantomsubfloat{\label{fig:intro_figure_output}}
        \phantomsubfloat{\label{fig:intro_figure_spring}}
        \phantomsubfloat{\label{fig:intro_figure_RLC}}
        \vspace{-4\baselineskip}
        \caption{An illustration of the model and an example of the training task.
        The model is a network (a) of bonds that can either represent damped springs (d) or RLC elements (e). A time-dependent signal is applied to a source node (b), and the network outputs (c) a desired response. In this example, the network classifies a sound signal---the spoken words ``zero'' or ``one''.
        }
        \label{fig:intro_figure}
    \end{figure*}

    Learning in the dynamical regime offers new opportunities.
    Mechanical systems could be trained to be frequency-dependent filters~\cite{ronellenfitsch2019inverse}, to transduce energy from one frequency to another~\cite{hexner2021training}, to perform acoustic protection~\cite{yang2008membrane} and locomotion~\cite{hermans2014automated}, and even to function as passive sensors~\cite{dubvcek2024sensor}.
    Unlike quasistatics, the time domain response depends on the integration over the history of past events.
    For example, one could imagine a passive electronic circuit that learns to classify sounds applied to an input node (\cref{fig:intro_figure}).

    In this work, we consider reciprocal systems (the response of node $i$ to a force on node $j$ is equal to the response of node $j$ to a force on node $i$)~\cite{kuo1966network,love2013treatise}.
    Reciprocal systems include any passive mechanical systems, such as spring networks (\cref{fig:intro_figure_spring}), and many electrical systems, such as RLC circuits (\cref{fig:intro_figure_RLC}).
    We consider linear systems and show that the dynamics extremize an effective action, enabling the use of \gls{eqprop} when the system operates under (i)~\gls{pbc} or (ii)~resting initial conditions.

    We begin with a brief background on \gls{eqprop} and introduce a generalized \gls{eqprop} framework for linear reciprocal dynamical systems.
    We then apply our algorithm to spring networks in periodic steady state and evaluate its robustness to nonlinearities.
    Finally, we demonstrate that RLC networks can be trained under both \gls{pbc} and resting initial conditions: in the former, to distinguish spoken digits, and in the latter, to distinguish between two distinct pulses.

    \section*{Background}
    \Acrfull{eqprop} is a method for computing the gradient of a cost function by exploiting the inherent physical laws.
    The condition for its validity is that the state of the system $\vector{x}$ extremizes an energy function $E$, which need not be the true energy.
    The goal is to compute the gradient of a cost function $C$ with respect to the learning degrees of freedom $\vector{\theta}$.
    The cost function measures the distance between the state of the system and the desired state.
    A perturbation is added to the energy, which acts as a force that nudges the system towards the desired state,
    \begin{equation}
        F = E + \beta C,
        \label{eq:total_energy}
    \end{equation}
    with $\beta$ being the nudging factor.

    The nudged state $\vector{x}^{(\beta)}$ extremizes the total energy $F$, under the assumed physical laws, with the added nudging forces.
    Finally, the gradient $\vector{\nabla}_{\vector{\theta}} C$ is given by~\cite{scellier2017equilibrium,laborieux2021scaling}.
    \begin{equation}
        \vector{\nabla}_{\vector{\theta}} C \approx \frac{1}{\beta}
        \left(
            \vector{\nabla}_{\vector{\theta}} F \left(\vector{x}^{(\beta / 2)}\right) -
            \vector{\nabla}_{\vector{\theta}} F \left(\vector{x}^{(-\beta / 2)}\right)
        \right).
        \label{eq:dCdtheta}
    \end{equation}
    Here, for higher precision, we use the central finite difference variation of \gls{eqprop}~\cite{laborieux2021scaling}.
    In the limit of $\beta\rightarrow0$, \cref{eq:dCdtheta} becomes exact.
    Thus, through two physical measurements, the gradient of the cost function is computed.

    \section*{Model}
    We develop an algorithm that enables the application of \gls{eqprop} to reciprocal systems operating in the dynamical regime in the linear limit.
    First, we focus on spring networks, which are governed by Newton's equations in the presence of damping.
    We then generalize our model and demonstrate that RLC networks can be treated similarly.
    Preparation details are provided in \refappendix{appendix:disordered_networks}.

    Consider a disordered network (\cref{fig:intro_figure_net}) of linear springs (\cref{fig:intro_figure_spring}) with potential energy $U = \frac{1}{2} \sum_j k_j \left( \ell_j - \ell_{j0} \right)^2$, $k_j$ being the spring constant, $\ell_j$ the length and $\ell_{j0}$ the rest length.
    If the displacements around the minimum are small, the forces can be linearized, yielding the following equations in vector notation:
    \begin{equation}
        M {\ddot{\vector{x}}} + \Gamma {\dot{\vector{x}}} + K \vector{x} = \vector{f}.
        \label{eq:linear_dynamics_spring}
    \end{equation}
    Here $\vector{x}$ is the vector of the node displacements, $\vector{f}$ the vector of external forces, $M_{kl}=m_k \delta_{kl}$ the mass matrix, $\Gamma_{kl}=\gamma_k \delta_{kl}$ the damping matrix, and $K = \vector{\nabla}_{\vector{x}}^2 U$ the Hessian --- the matrix of second derivatives of the potential energy at equilibrium.\footnote{$m_k$ ($\gamma_k$) is the mass (damping) associated with the node to which the $k$\textsuperscript{th} coordinate belongs.}
    Note that the matrices $M$, $\Gamma$, and $K$ are symmetric, which is a consequence of the reciprocity of the system.

    This particular form of the dissipation term is chosen for simplicity and can easily be extended to other models, such as the Kelvin-Voigt model (where $\Gamma$ is no longer diagonal, but remains symmetric).
    Note that, in general, a network of linear springs exhibits nonlinear behavior at large displacements due to changes in the angles between elements.

    It is convenient to recast the equations of motion (\ref{eq:linear_dynamics_spring}) using the differential operator $H$:
    \begin{equation}
        H \vector{x} \triangleq \left[M \frac{\D^2}{\D t^2} + \Gamma \frac{\D}{\D t} + K\right] \vector{x} =  \vector{f}.
        \label{eq:linear_dynamics}
    \end{equation}
    While we have focused on elastic networks, $H$ can represent an arbitrary reciprocal linear dynamical system.
    For example, a disordered circuit composed of RLC blocks, shown in \cref{fig:intro_figure_RLC}, can be described similarly.
    In this circuit, every node $i$ is connected to a common ground by a resistor $R_i = 1 / g_i$ and a capacitor $C_i$ in parallel.
    Inductors $L_{ij}$ are placed between nodes $i$ and $j$.
    Such an RLC network is analogous to a spring network, where the resistors, capacitors, and inductors are related to the damping, mass, and spring constants, respectively.
    This network can be described using \cref{eq:linear_dynamics}, with $\vector{x}$ representing the node voltages, and $\vector{f}$ representing the external rate of change of the external currents applied to the nodes.
    To this end, the matrices $M$, $\Gamma$, and $K$ should be replaced by the capacitance matrix, the conductance matrix, and the matrix of inverse inductances, respectively, as detailed in \refappendix{appendix:RLC_equations}.
    Of course, such an RLC circuit is only one example, and other topologies can be studied as well.

    \section*{Algorithm}

    \subsection*{An effective action for dissipative systems}
    \gls{eqprop} requires that the dynamics extremize an ``energy'' function.
    When there is no dissipation, a natural candidate is the action.
    We present an effective dissipative action, whose extremum corresponds to trajectories defined by the equations of motion.
    Under \gls{pbc}, i.e.\ $\vector{x}(t + \period) = \vector{x}(t)$, the effective action is most easily derived in Fourier space, where the variation corresponds to extremizing with respect to the Fourier coefficients.
    This defines a unique effective action (up to unimportant multiplicative factors), which we present in the time domain:
    \begin{equation}
        E = \frac{1}{\period} \int_{-\period/2} ^{\period/2} \left[ \frac{1}{2}  \vector{x}(-t)^\top H \vector{x}(t) - \vector{x}(-t)^\top \vector{f}(t)\right] \D t.
        \label{eq:energy_function}
    \end{equation}
    It is straightforward to verify, as detailed in \refappendix{appendix:derivation_action}, that $E$ is stationary under (i) periodic variations of the steady state, and (ii) resting initial conditions.
    We make this distinction to emphasize that for resting initial conditions, no driving occurs prior to inference (or training).
    This setting is particularly convenient as it provides a natural initial condition for inference for passive sensors.

    \Cref{eq:energy_function} closely resembles the extended action for dissipative systems, formulated using an auxiliary field that obeys time-reversal dynamics~\cite{bateman1931dissipative,morse1953methods,feshbach1977quantization,celeghini19891,celeghini1992quantum,galley2013classical}.
    Note that the effective action depends on the time-reversed trajectory and, therefore, the integrand (effective Lagrangian) is non-local in time.

    Lagrangian dynamics in the linear regime extremizes both the action and the effective action, and therefore, one could imagine that they are related.
    Indeed, \cref{eq:energy_function} is similar to the action in the limit of vanishing damping, with one key difference.
    The action includes only forward-in-time trajectories; replacing $\vector{x}(-t)$ by $\vector{x}(t)$ in the effective action yields the action.
    Even in the absence of dissipation, the action and the effective action are, in general, different.

    \subsection*{The nudging forces}
    Now that we have demonstrated the suitability of the effective action, we can employ \gls{eqprop}.
    A subset of the nodes is designated as the target set (denoted by $\mathrm{T}$), and our goal is to realize a desired periodic motion $\vector{x}_\mathrm{D}(t)$.
    Next, we define a cost function,
    \begin{equation}
        C \triangleq \frac{1}{\period} \int_{-\period/2} ^{\period/2}  \left\| \vector{x}_\mathrm{T}(t) - \vector{x}_\mathrm{D}(t) \right\|^2 \D t
        \label{eq:cost_function}
    \end{equation}
    where the 2-norm is chosen for simplicity.
    We compute the nudged dynamics by extremizing the total effective action defined in \cref{eq:total_energy}, with the derivation provided in \refappendix{appendix:nudging}:
    \begin{equation}
        H \vector{x}^{(\beta)} = \vector{f} + 2 \beta \delta_\mathrm{T} \mathcal{T} \left(\vector{x}_\mathrm{D} - \vector{x}_\mathrm{T}^{(\beta)} \right),
        \label{eq:nudged_dynamics_beta}
    \end{equation}
    where $\mathcal{T}$ denotes the time-reversal operator and $\delta_\mathrm{T}$ is an indicator function equal to unity for targets and zero otherwise; thus, only target nodes are nudged.

    The presence of $\mathcal{T} \vector{x}_\mathrm{T}^{(\beta)}$ in \cref{eq:nudged_dynamics_beta} is concerning since the forward-in-time solution depends on the time-reversed solution.
    To address this, we replace $\vector{x}_\mathrm{T}^{(\beta)}$ on the right-hand side of \cref{eq:nudged_dynamics_beta} with $\vector{x}_\mathrm{T}^{(0)}$, corresponding to the free (unnudged) dynamics.
    Since this offers a sub-leading correction of order $\mathcal{O}(\beta)$, the algorithm still converges for $\beta \rightarrow 0$.
    Thus, we obtain
    \begin{equation}
        H \vector{x}^{(\beta)} = \vector{f} + 2 \beta \delta_\mathrm{T} \mathcal{T} \left(\vector{x}_\mathrm{D} - \vector{x}_\mathrm{T}^{(0)} \right).
        \label{eq:nudged_dynamics_zero}
    \end{equation}
    While this requires additional computations compared to the quasistatic case, time reversal is applied only to the targets and can therefore be performed externally.
    Additionally, time reversal (and measurement) need only be done for $|t| < \period/2$, limiting the required data to that interval.

    In summary, learning is implemented as follows.
    (a)~Record the free trajectory $\vector{x}^{(0)}$.
    (b)~Compute the required nudging forces.
    (c)~Record the nudged trajectories $\vector{x}^{(\pm \beta / 2)}$.
    (d)~Calculate the gradient $\vector{\nabla}_{\vector{\theta}} C$ using \cref{eq:dCdtheta}.
    (e)~Update the parameters $\vector{\theta}$ using gradient descent (or another optimizer) and repeat.

    It is important to note that changing the parameters or nudging perturbs the trajectory of the system.
    In the case of \gls{pbc}, one must allow the system to relax to a new steady state.
    Physical parameters, particularly damping, govern the number of relaxation periods~\cite{berneman2024designing}.
    For resting initial conditions, one must allow the system to relax before proceeding to the next cycle.
    In our simulations, we made sure that these conditions were met.

    \subsection*{Update rules for specific systems}
    Within \gls{eqprop}, the gradient of the cost function is computed using \(\vector{\nabla}_{\vector{\theta}} F\) (see \cref{eq:dCdtheta}).
    We provide explicit formulas for this quantity for both spring and RLC networks.
    Notably, the resulting update rules are identical under both \gls{pbc} and resting initial conditions.

    \subsubsection*{Spring networks}
    First, we consider the case where the damping coefficients are the learning parameters ($\theta_i = \gamma_i$),
    \begin{equation}
        \frac{\partial F}{\partial \gamma_i} = \frac{1}{2 \period} \int_{-\period/2} ^{\period/2} \vector{r}_i(-t)^\top \dot{\vector{r}}_i(t) \D t,
        \label{eq:dFdgamma}
    \end{equation}
    with $\vector{r}_i$ being the steady state displacement of the $i$\textsuperscript{th} node.
    Thus, the update rule requires additional memory and computational resources compared to the quasistatic case.

    When the spring constants are the learning parameters ($\theta_j = k_j$),
    \begin{equation}
        \frac{\partial F}{\partial k_j} = \frac{1}{2 \period} \int_{-\period/2} ^{\period/2} \ell_{j \parallel}(-t) \ell_{j \parallel}(t) \D t,
        \label{eq:dFdk}
    \end{equation}
    with $\ell_{j \parallel}$ being the displacement of the $j$\textsuperscript{th} spring parallel to the spring at equilibrium.
    In this derivation, we have focused on networks whose equilibrium is unstressed~\cite{lubensky2015phonons,hagh2022rigidpy}.
    If stresses are present, the equilibrium positions change, further complicating the problem.
    This also occurs when the rest lengths are varied.

    \subsubsection*{RLC networks}
    For RLC networks, similar arguments apply.
    When the conductances (inverse resistances) are the learning parameters ($\theta_i = g_i$),
    \begin{equation}
        \frac{\partial F}{\partial g_i} = \frac{1}{2 \period} \int_{-\period/2} ^{\period/2} V_i(-t) \dot{V}_i(t) \D t.
        \label{eq:dFdg}
    \end{equation}
    with $V_{i}$ being the voltage at the $i$\textsuperscript{th} node.
    Similar rules could also be derived for the capacitors and inductors.

    Developing approximate schemes could allow for efficient and practical physical implementation.
    For example, in the case of \gls{pbc}, if one trains one frequency at a time, the most important information is the sign of each term.
    The sign depends solely on the phase relative to the driving, $\phi$; for example, $ \mathrm{sign}\left(\frac{\partial F}{\partial g_i}\right)= \mathrm{sign}(\sin (2\phi))$.
    Therefore, it is sufficient to estimate the phase---for example, by comparing the time difference between the maxima of the node voltage and the driving.

    \section*{Simulations}
    We apply our algorithm to two different systems: spring and RLC networks.
    We first consider \gls{pbc} in the context of both networks and then demonstrate the algorithm for resting initial conditions in RLC networks.
    Simulation details are provided in \refappendix{appendix:simulations}.

    \subsection*{Spring networks under \gls{pbc}}
    We consider an ensemble of disordered spring networks with 50 nodes.
    As noted, spring networks are nonlinear in general, allowing us to test the effect of nonlinear dynamics in our linear \gls{eqprop} method.
    The nonlinear contribution grows with the strain amplitude of the deformation and is subleading for small deformations.

    A single source node is driven sinusoidally in the $x$-direction with amplitude $A$, i.e.\ $x_\mathrm{S}(t) = A \cos{(\Omega t)}$.
    The goal, chosen for simplicity, is for a randomly chosen target node to move $90^\circ$ out of phase from the source node with equal amplitude, i.e.\ $x_\mathrm{D}(t) = A \sin{(\Omega t)}$.
    To achieve this goal, we train the network using gradient descent, where \gls{eqprop} is used to calculate the gradient of the cost function.\footnote{We use a variant of the cost function $C_\mathrm{norm} = C / A^2$ that is independent of the amplitude in the limit $A \rightarrow 0$.}
    For reference, we also compute the gradient with \gls{backprop} (see~\cite{berneman2024designing} for details), which does not require the assumption of linearity.

    We perform two sets of experiments: in the first set, we modify the damping coefficients $\gamma_i$ of the nodes; in the second set, we modify the spring constants $k_j$.
    More details are provided in \refappendix{appendix:spring_network_details}.

    \Cref{fig:opt_vs_eqprop_different_amplitudes1,fig:opt_vs_eqprop_different_amplitudes2} show the median error as a function of the epochs for systems where the damping coefficients and spring constants are adapted, respectively.
    In both cases, \gls{backprop} outperforms \gls{eqprop}.
    At small amplitudes, the two methods are only distinguishable at later epochs.
    As the amplitude increases, the difference between the two methods becomes more pronounced.

    \begin{figure}[!ht]
        \centering
        \subfloat{
            \begin{overpic}{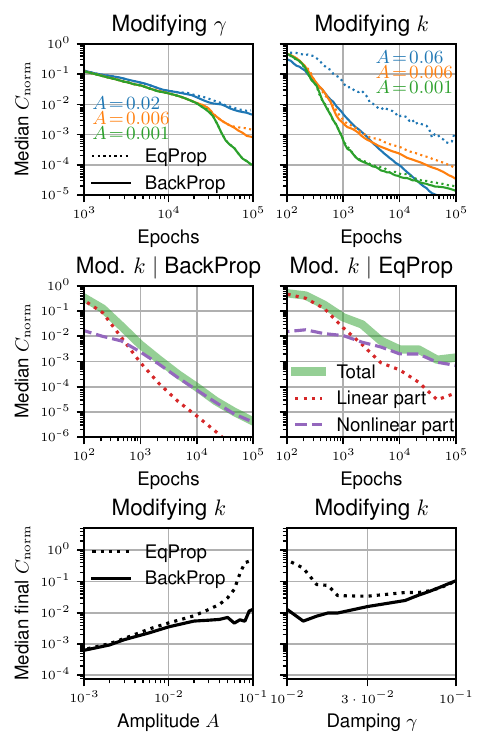}
                \put(5.5,96){\mybox{a}}
                \put(33.5,96){\mybox{b}}
                \put(5.5,66){\mybox{c}}
                \put(33.5,66){\mybox{d}}
                \put(5.5,31){\mybox{e}}
                \put(33.5,31){\mybox{f}}
            \end{overpic}
            \label{fig:opt_vs_eqprop_different_amplitudes1}
        }
        \phantomsubfloat{\label{fig:opt_vs_eqprop_different_amplitudes2}}
        \phantomsubfloat{\label{fig:opt_vs_eqprop_linear_and_nonlinear_error1}}
        \phantomsubfloat{\label{fig:opt_vs_eqprop_linear_and_nonlinear_error2}}
        \phantomsubfloat{\label{fig:opt_vs_eqprop_amplitudes}}
        \phantomsubfloat{\label{fig:opt_vs_eqprop_gammas}}
        \vspace{-3\baselineskip}
        \caption{Training spring networks to assess the effect of nonlinearities.
        A sinusoidal displacement is applied to a source to attain sinusoidal motion on the target with a phase difference of $90^\circ$.
            (a),(b)~The median error $C_\mathrm{norm}$ as a function of the epochs using \gls{eqprop} and \gls{backprop}, for different amplitudes $A$. In (a), we modify the damping coefficients $\gamma_i$ and in (b) the spring constants $k_j$.
            In (c) and (d), we decompose the error into linear and nonlinear contributions. Note that \gls{eqprop} fares worse in reducing the nonlinear contribution.
            (e),(f) Comparison of the error for \gls{eqprop} and \gls{backprop} after 1000 epochs, as a function of the amplitude (e) and the damping coefficients (f).
            Parameters: $m = 1$, $k_\mathrm{init} = 1$, $\beta = 10^{-9}$, $\Omega=0.5$, learning rate~=~0.01, 100 periods to reach steady state. (a) $\gamma_\mathrm{init} = 0.1$. (b),(c),(d),(e) $\gamma_\mathrm{init} = 0.01$. (c),(d) $A = 0.06$.
        }
    \end{figure}
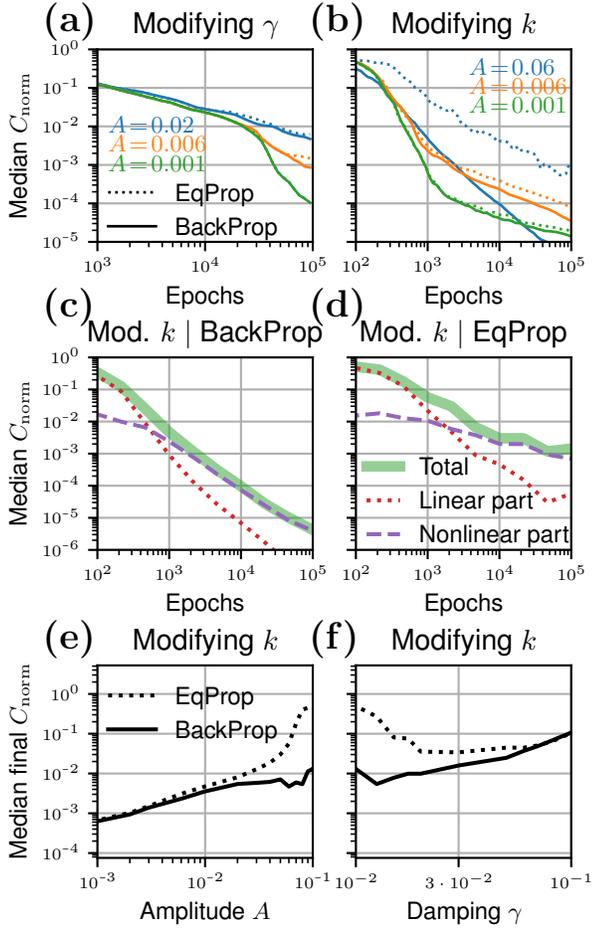

    To gain insight into the effect of nonlinearities on \gls{eqprop} (especially for $A = 0.06$ in \cref{fig:opt_vs_eqprop_different_amplitudes2}), we break the error down into its linear and nonlinear contributions~\cite{berneman2024designing}.
    We associate the Fourier component at the driving frequency $\Omega$ with the linear contribution to the error and the remaining with the nonlinear contribution.
    \Cref{fig:opt_vs_eqprop_linear_and_nonlinear_error1,fig:opt_vs_eqprop_linear_and_nonlinear_error2} show the median error as a function of the epochs, broken down to the linear and nonlinear terms for $A = 0.06$.
    For both \gls{eqprop} and \gls{backprop}, the nonlinear contribution is initially smaller and decreases at a slower rate, up to a point where it dominates the total error.
    In the case of \gls{backprop}, shown in \cref{fig:opt_vs_eqprop_linear_and_nonlinear_error1}, the error decreases continually, even when the nonlinearities dominate the error.
    In contrast, as shown in \cref{fig:opt_vs_eqprop_linear_and_nonlinear_error2}, \gls{eqprop} reduces the error at a slower pace, where most of the decrease in the error can be attributed to a reduction of the linear contribution.

    Finally, we compare the performance of \gls{eqprop} and \gls{backprop} for different amplitudes $A$ and damping coefficients $\gamma$ in \cref{fig:opt_vs_eqprop_amplitudes,fig:opt_vs_eqprop_gammas}, respectively.
    \Gls{eqprop} performs as well as \gls{backprop} for small amplitudes (large damping), and underperforms as the amplitude increases (damping decreases) due to the increasing presence of nonlinearities.
    In summary, nonlinearities degrade performance gradually, allowing \gls{eqprop} to remain effective at small---but finite---amplitudes.

    \subsection*{RLC networks}

    Here we demonstrate that a 50-node RLC network (\cref{fig:intro_figure_net,fig:intro_figure_RLC}) can be trained to classify temporal sequences by tuning the conductances (inverse resistances).
    We consider two scenarios: (i)~\gls{pbc}, where the system is driven by a periodic signal, and (ii)~resting initial conditions, where the system is trained to classify pulses.

    \subsubsection*{\Acrfull{pbc}}
    We train an RLC network to differentiate between recordings of spoken words.
    We consider the audible digits taken from the audioMNIST dataset~\cite{audiomnist2023}; for simplicity, we focus on the first two digits (``zero'' and ``one''), as illustrated in \cref{fig:intro_figure_input,fig:intro_figure_output}.
    We follow the data splitting and preprocessing steps for the AudioNet model described in~\cite{audiomnist2023}.
    In addition, we discard the DC component of the audio signal.
    Here, preprocessing is minimal; the circuit is fed with time-domain data.

    We randomly select a source node $\mathrm{S}$ and two target bonds $\mathrm{T}_0, \mathrm{T}_1$.
    Data is applied to the source node by varying the current $I_\mathrm{S}$, while the target bonds are nudged with voltages $V_{\mathrm{nudge},\mathrm{T}_0}, V_{\mathrm{nudge},\mathrm{T}_1}$.
    To classify the signal, the sound snippet is applied periodically to the source until a steady state is reached.
    The predicted digit is deduced from the signal energy (SE) on the target bonds:
    \begin{equation}
        \mathrm{SE}_{\mathrm{T}_i} = \frac{1}{\period} \int_{-\period/2} ^{\period/2} \left\| V_{\mathrm{T}_i}(t) \right\|^2 \D t, \ i = 0,1.
        \label{eq:SE}
    \end{equation}
    If the signal energy at $\mathrm{T}_0$ is greater than the signal energy at $\mathrm{T}_1$, we predict the digit to be ``zero''; otherwise, we predict it to be ``one''.
    Given the binary nature of the problem, we use the softmax function to calculate the probability of the prediction, and the cross-entropy loss function to calculate the cost~\cite{lecun2015deep}.
    More details on the cost, nudging voltages, \gls{eqprop} rule, and training procedure are given in~\refappendix{appendix:RLC_training}.

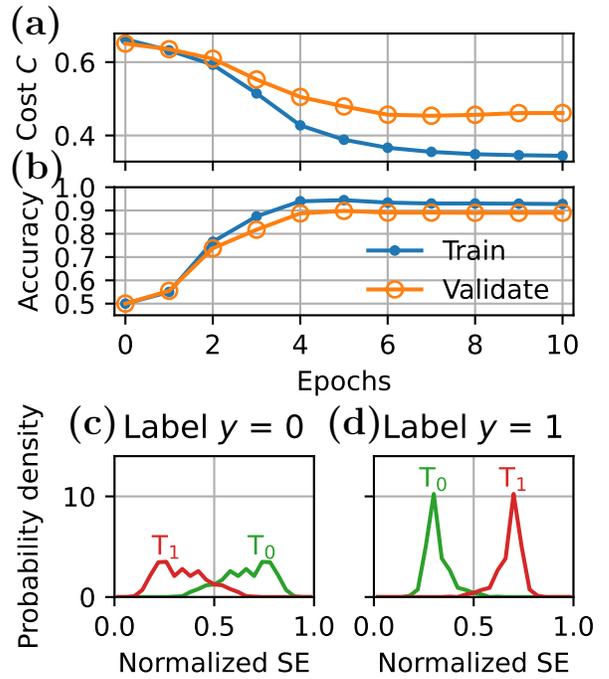
\begin{figure}[!ht]
        \centering
        \subfloat{
            \begin{overpic}{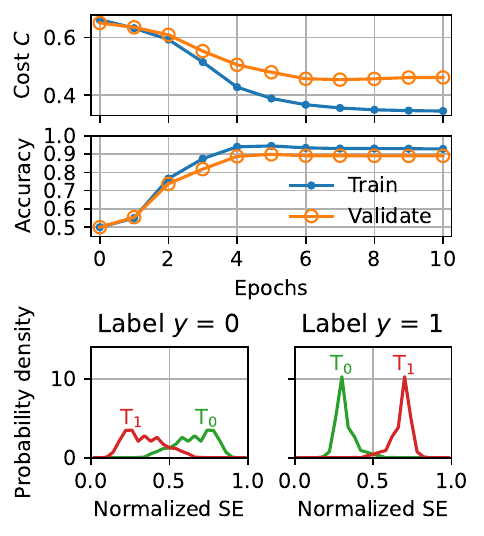}
                \put(1.5, 97){\mybox{a}}
                \put(1.5, 76){\mybox{b}}
                \put(10, 38.5){\mybox{c}}
                \put(48, 38.5){\mybox{d}}
            \end{overpic}
            \label{fig:MNIST_error}
        } \\
        \phantomsubfloat{\label{fig:MNIST_accuracy}}
        \phantomsubfloat{\label{fig:MNIST_SE1}}
        \phantomsubfloat{\label{fig:MNIST_SE2}}
        \vspace{-3\baselineskip}
        \caption{RLC network operating in periodic steady state, trained on the audioMNIST dataset to classify the spoken digits ``zero'' and ``one''.
            (a)~The error $C$ for the training and validation datasets. (b)~The accuracy (percentage of correctly classified inputs) for the training and validation datasets.
            (c),(d)~Probability densities of the normalized signal energy (SE) at the target bonds used to differentiate between ``zero'' (c) and ``one'' (d) from the test dataset.
        }
    \end{figure}

    The error and accuracy for the training dataset and validation dataset (which is not used for training) are shown in \cref{fig:MNIST_error,fig:MNIST_accuracy}, respectively.
    At the final epoch, the RLC network achieves a test accuracy of 92.8\%.
    Thus, we achieve high precision without carefully tuning the topology or using large systems.

    The probability densities of the normalized signal energy $\mathrm{SE}_{\mathrm{T}_i} / (\mathrm{SE}_{\mathrm{T}_0} + \mathrm{SE}_{\mathrm{T}_1})$ for test data labeled ``zero'' and ``one'' are shown in \cref{fig:MNIST_SE1,fig:MNIST_SE2}, respectively.
    There is some overlap in the distributions, which results in misclassification.
    This is more significant for data labeled ``zero'' (\cref{fig:MNIST_SE1}).

    In summary, we have demonstrated that simple electrical networks can be used to perform temporal computations and to solve classification problems.
    Presumably, mechanical systems can perform similar tasks, which could be used as passive sensors driven by sound signals.

    \subsubsection*{Resting initial conditions}
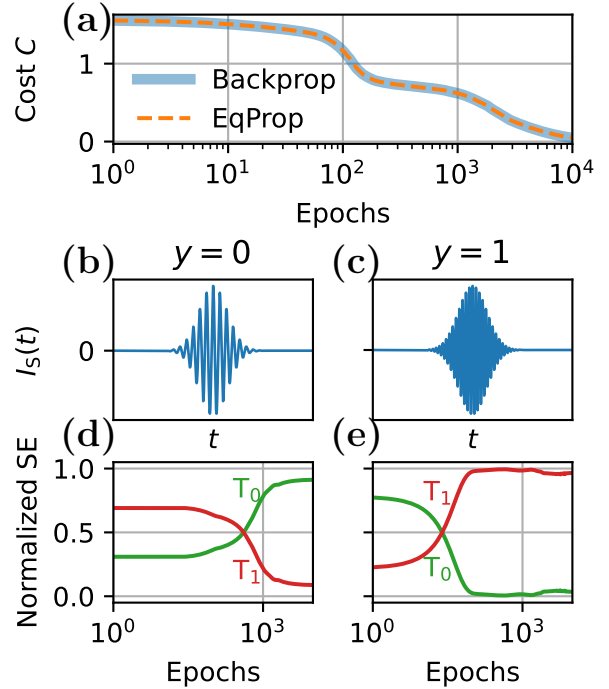
\begin{figure}[!ht]
        \centering
        \subfloat{
            \begin{overpic}{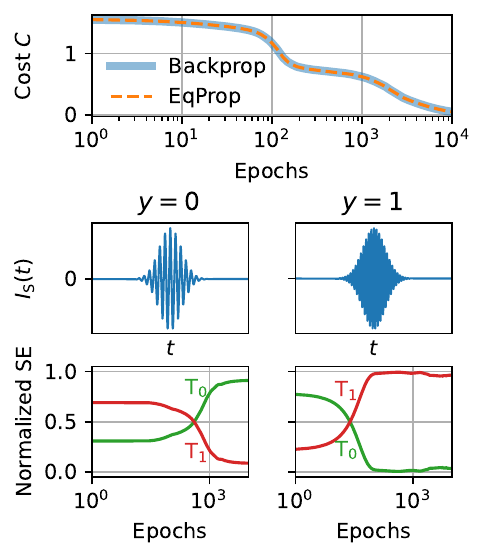}
                \put(9, 95){\mybox{a}}
                \put(9, 61){\mybox{b}}
                \put(47, 61){\mybox{c}}
                \put(9, 36.5){\mybox{d}}
                \put(47, 36.5){\mybox{e}}
            \end{overpic}
            \label{fig:BC_error}
        } \\
        \phantomsubfloat{\label{fig:BC_Iext1}}
        \phantomsubfloat{\label{fig:BC_Iext2}}
        \phantomsubfloat{\label{fig:BC_SE1}}
        \phantomsubfloat{\label{fig:BC_SE2}}
        \vspace{-3\baselineskip}
        \caption{RLC network with resting initial conditions, trained to distinguish between two distinct pulses.
            (a)~Error $C$, with the gradient computed using \gls{eqprop} and compared against \gls{backprop} as a reference.
            (b),(c)~The external current $I_\mathrm{S}$ applied to the source node, for the two pulses used in the training.
            (d),(e)~The normalized signal energies (SE) at the target bonds used to differentiate between the two pulses.
        }
    \end{figure}

    Using the same RLC network and training procedure, we now demonstrate that \gls{eqprop} can be used to train networks with resting initial conditions.
    We apply either one of two pulses, shown in \cref{fig:BC_Iext1,fig:BC_Iext2}, to the source node, and the goal is to classify the pulse based on the signal energy (\cref{eq:SE}) at the target bonds.
    The two pulses have the same Gaussian envelope, but one operates at $f_\mathrm{S} = 600 \ \mathrm{Hz}$, while the other operates at $f_\mathrm{S} = 1200 \ \mathrm{Hz}$:
    \begin{equation}
        I_\mathrm{S}(t) = \E^{-\frac{1}{2} \left(\frac{t}{\sigma}\right)^2} \cos{(2 \pi f_\mathrm{S} t)}.
    \end{equation}
    with $\sigma = 5 \ \mathrm{ms}$.
    We simulate the circuit for $\period = 50 \ \mathrm{ms}$, with the system being at rest at $t = -\period / 2$.

    We use \gls{eqprop} to train the network to classify the two pulses, and use \gls{backprop} for comparison.
    \Cref{fig:BC_error} shows the error $C$ as a function of the epochs, clearly demonstrating that \gls{eqprop} is equivalent to \gls{backprop} in this case.
    \Cref{fig:BC_SE1,fig:BC_SE2} show the normalized signal energy at the target bonds for the two pulses.
    By the end of training, $\mathrm{SE}_{\mathrm{T}_0}$ dominates for the first pulse (\cref{fig:BC_Iext1,fig:BC_SE1}), while $\mathrm{SE}_{\mathrm{T}_1}$ dominates for the second pulse (\cref{fig:BC_Iext2,fig:BC_SE2}).

    \section*{Conclusion}
    We have shown that \acrfull{eqprop} extends to linear reciprocal dynamical systems operating either in periodic steady state or with resting initial conditions, even in the presence of dissipation.
    In this context, we have introduced an effective action, which is related to the action for undamped systems and can be interpreted as a generalization of the action to dissipative systems.
    This extension to dissipative systems is useful because many systems inevitably contain some dissipation, whether through friction in mechanical systems or through parasitic resistances in electrical systems.

    \Gls{eqprop} in the dynamical regime introduces new challenges.
    The update rule requires integrating over the time-dependent trajectory, its time-reversed counterpart, and possibly its derivatives.
    While this can be done using electronics, it requires both memory and computing resources.
    Training one frequency at a time may allow for simplifying these calculations since they only depend on the relative phase between the response and the forcing.
    Perhaps, approximate update rules can be developed that could allow for the efficient implementation of these methods.

    The application of \gls{eqprop} to systems under \acrfull{pbc} implies that a steady state must be maintained both during training and inference.
    For sensors, this means that the input signal must be applied repeatedly until a steady state is reached during inference.
    In contrast, the case of resting initial conditions is particularly appealing, since inference can be done from a single wave packet. Furthermore, for dissipative dynamics, the system naturally relaxes to rest.

    Using the framework we developed, we have successfully trained spring and RLC networks to perform desired tasks.
    In the case of spring networks, \gls{eqprop} computes the gradient of the cost function exactly for weak driving, in the linear regime.
    Its effectiveness diminishes as the driving amplitude increases due to nonlinear contributions.
    Nevertheless, these nonlinear effects appear to be perturbative in the amplitude, and for finite and small driving amplitudes, training is still successful.

    We have also explored the ability to perform complex tasks such as classification.
    The dynamical regime allows for responses that depend on the whole range of frequencies.
    This allowed us, for example, to classify audible signals with high accuracy (digits ``zero'' and ``one'') using an RLC network operating in periodic steady state.
    We have also demonstrated that the same RLC network with resting initial conditions can be trained to classify non-repeating pulses.

    This work opens the door to self-learning systems in the dynamic regime.
    Harnessing dynamics allows for responses that are not compatible with quasistatics, such as frequency-dependent responses, responses at an arbitrary phase, and responses that are dependent on the temporal sequence of the driving.

    Our results show that \gls{eqprop} applies to arbitrary linear reciprocal dynamical systems.
    In the undamped nonlinear regime, a rule based on the action can also be employed.
    Currently, the damped nonlinear regime remains inaccessible.
    Perhaps, interpolating between the two methods may provide access to this regime.

    \section*{Acknowledgments}
    We would like to acknowledge Maor Eldar for insightful discussions.
    In addition, we would like to thank Serge Massar for his comments on the manuscript.
    This work was supported by the Israel Science Foundation (grant 2385/20) and the Alon Fellowship.

    \printbibliography
    \appendix

    \section{Disordered networks}
    \label{appendix:disordered_networks}
    We generate disordered networks from packings of polydisperse soft spheres~\cite{o2003jamming}.
    The packings are prepared with 50 particles, prior to the removal of rattlers, with a coordination number $Z \approx 4.42$.
    For spring networks, overlapping spheres are connected by linear springs, with rest lengths set to the original distance of the springs to avoid prestress.
    A source, target, and 2 fixed nodes are assigned randomly, where care is taken to make sure that a bond does not connect these nodes.
    For RLC networks, inductors connect the nodes, and each node is connected to a common ground by a resistor and a capacitor in parallel.
    Similarly, a source node and two target bonds are assigned randomly.

    \section{RLC network equations}
    \label{appendix:RLC_equations}
    The equations of motion for the RLC network shown in \cref{fig:intro_figure_net,fig:intro_figure_RLC} are given by
    \begin{align}
        C \frac{\D \vector{V}}{\D t} + G \vector{V} &= \vector{I}_\mathrm{S} + B \vector{i}_L, \label{eq:RLC_V} \\
        L \frac{\D \vector{i}_L}{\D t} + B^\top \vector{V} &= 0, \label{eq:RLC_I}
    \end{align}
    with $C$, $G$, and $L$ being diagonal capacitance, conductance (inverse resistance), and inductance matrices, respectively, $\vector{V}$ being the node voltages, $\vector{i}_L$ being the inductor currents, and $\vector{I}_\mathrm{S}$ being external node currents.
    $B$ is the incidence matrix, with $B_{ij} = \pm 1$ if the $j$\textsuperscript{th} inductor is attached to the $i$\textsuperscript{th} node, with the sign indicating the direction of the current, and $B_{ij} = 0$ otherwise.
    Deriving \cref{eq:RLC_V} with respect to time, and plugging in \cref{eq:RLC_I}, we obtain
    \begin{equation}
        C \ddot{\vector{V}} + G \dot{\vector{V}} + K \vector{V} = \dot{\vector{I}}_\mathrm{S},
        \label{eq:RLC_equation}
    \end{equation}
    with $K = B L^{-1} B^\top$.
    Comparing \cref{eq:RLC_equation} to \cref{eq:linear_dynamics}, we can write $H = C \frac{\D^2}{\D t^2} + G \frac{\D}{\D t} + K$, $\vector{x} = \vector{V}$ and $\vector{f} = \dot{\vector{I}}_\mathrm{S}$.

    \section{Effective action}
    \label{appendix:derivation_action}
    We show that $E$, given by \cref{eq:energy_function}, is stationary under variations of the trajectory $\delta \vector{x}(t)$ under \gls{pbc} and resting initial conditions.
    \begin{align}
        \delta E = \frac{1}{\period} \int_{-\period/2} ^{\period/2} \Big[ & \frac{1}{2} \delta \vector{x}(-t)^\top H \vector{x}(t) + \frac{1}{2} \vector{x}(-t)^\top H \delta \vector{x}(t)) \nonumber \\
        & - \delta \vector{x}(-t)^\top \vector{f}(t) \Big] \D t
    \end{align}
    Using integration by parts on the second term, and assuming that $H$ is a reciprocal operator, we can write
    \begin{gather}
        \delta E = \frac{1}{\period} \int_{-\period/2} ^{\period/2} \left[\delta \vector{x}(-t)^\top (H \vector{x}(t) - \vector{f}(t)) \right] \D t \nonumber \\
        + \left[\text{boundary terms}\right]_{-\period/2}^{\period/2}.
    \end{gather}
    Given that $H \vector{x} = \vector{f}$ (see \cref{eq:linear_dynamics}), the first term is zero.
    Assuming the boundary terms vanish, we get $\delta E = 0$, which is what we need to be able to apply \gls{eqprop}.
    There are two interesting cases under which the boundary terms vanish:
    \begin{itemize}
        \item \Acrfull{pbc}, i.e.\ $\vector{x}(-\period/2) = \vector{x}(\period/2)$ and $\delta \vector{x}(-\period/2) = \delta \vector{x}(\period/2)$.
        \item Resting initial conditions, i.e.~the $k$\textsuperscript{th} derivative \ $\vector{x}^{(k)}(-\period/2) = \delta \vector{x}^{(k)}(-\period/2)  = 0$, where $0 \leq k < n$, $k \in \mathbb{N}$, and $n$ being the order of the differential operator $H$.
    \end{itemize}

    \section{Nudging term}
    \label{appendix:nudging}
    Introducing $C$ into the total effective action $F$ (see \cref{eq:total_energy,eq:cost_function}) yields a nudging force, which can be computed from the variation:
    \begin{align}
        \beta \delta C &= \frac{1}{\period} \int_{-\period/2} ^{\period/2} \delta \vector{x}_\mathrm{T}(t)^\top \left[2 \beta \left(\vector{x}_\mathrm{T}(t) - \vector{x}_\mathrm{D}(t)\right)\right] \D t \nonumber \\
        &= \frac{1}{\period} \int_{-\period/2} ^{\period/2} \delta \vector{x}_\mathrm{T}(-t)^\top \left[2 \beta \left(\vector{x}_\mathrm{T}(-t) - \vector{x}_\mathrm{D}(-t)\right)\right] \D t.
    \end{align}
    The expression in the parentheses corresponds to the nudging in \cref{eq:nudged_dynamics_beta}.

    \section{Simulation details}
    \label{appendix:simulations}
    We use the Embedded Runge-Kutta method of order 5(4)~\cite{tsitouras2011runge} to integrate the dynamics of the system.
    Integration is facilitated using JAX~\cite{jax} and Diffrax~\cite{diffrax}.
    In addition, the forces in the spring networks are calculated using the JAX MD package~\cite{jaxmd}.
    Specifically, when simulating the RLC network with periodic source currents, the steady state motion is obtained directly, without the need for integration.

    \section{Training the spring network}
    \label{appendix:spring_network_details}
    The spring constants and damping coefficients are initialized to the same value.
    When adapting the spring constants $k_j$, we limit their change to the range $[0.1, 10]$ to avoid any numerical instabilities.
    The only limit placed on the damping coefficients $\gamma_i$ is that they must be positive.

    \section{Training the RLC network}
    \label{appendix:RLC_training}
    Upon defining a cost function, the nudged dynamics will have the following form:
    \begin{equation}
        C \ddot{\vector{V}} + G \dot{\vector{V}} + K \vector{V} = \dot{\vector{I}}_\mathrm{S} + \vector{F}_\mathrm{nudge}
        \label{eq:nudged_dynamics_RLC}
    \end{equation}
    Assuming the cost function relates to voltages over bonds (inductors), we can write the nudging force as $\vector{F}_\mathrm{nudge} = B \vector{f}_\mathrm{nudge}$.
    Going back to a system of first-order differential equations, we find that only \cref{eq:RLC_I} must be adapted to account for nudging:
    \begin{equation}
        L \frac{\D \vector{i}_L}{\D t} + B^\top \vector{V} = L \vector{f}_\mathrm{nudge}.
    \end{equation}
    Thus, nudging can be implemented by adding a voltage source in series with the inductors, with voltage $\vector{V}_\mathrm{nudge} = L \vector{f}_\mathrm{nudge}(t)$.

    The signal energy (SE) at two target bonds (inductors) $\mathrm{T}_0$ and $\mathrm{T}_1$ is used to classify the input signal.
    The probability of predicting ``one'' is calculated using the softmax function:
    \begin{equation}
        p_1 = \frac{\E^{\mathrm{SE}_{\mathrm{T}_1}}}{\E^{\mathrm{SE}_{\mathrm{T}_0}} + \E^{\mathrm{SE}_{\mathrm{T}_1}}}.
    \end{equation}
    This quantity is used to calculate the cross-entropy loss function,
    \begin{equation}
        C = - \left[ y \log{p_1} + (1 - y) \log{(1 - p_1)} \right],
    \end{equation}
    with $y$ indicating the true label ($y = 0$ for ``zero'', $y = 1$ for ``one'').

    Defining the effective action using \cref{eq:energy_function} and the total effective action using \cref{eq:total_energy} gives the following nudging voltages:
    \begin{equation}
        V_{\mathrm{nudge},\mathrm{T}_i} = 2 \beta L_{\mathrm{T}_i} \cdot (-1)^{y + i} \left[ (1 - y) p_1 + y (1 - p_1) \right] V_{\mathrm{T}_i}(-t),
    \end{equation}
    with $i = 0,1$.
    The gradient is obtained using \cref{eq:dCdtheta,eq:dFdg}.

    We use the Adam optimizer~\cite{kingma2014adam}, with an initial learning rate of $10^{-4}$.
    To avoid instabilities, we clip the conductances such that $R_i = g_i^{-1} \leq 1 \ \si{\mega \ohm}$.
    The conductances are all initialized to $100 \ \si{\per\ohm}$, the capacitances are fixed to $10 \ \si{\micro \farad}$, and the inductances to $5 \ \si{\milli \henry}$, ensuring that the eigenfrequencies of the network are of order $1 \ \si{\kilo \hertz}$, which lies in the frequency range of interest for audible signals.

    \paragraph{AudioMNIST task}
    For the audioMNIST task, we update the conductances in batches of 120 training samples.
    In total, we run the training procedure for 10 epochs, with each epoch being defined as a full pass over the training dataset.

\end{document}